# EVIDENCE OF CURRENT STABILIZATION AFTER A LONG-TIME DECAY IN HIGH-$T_C$ SUPERCONDUCTORS

H. González-Jorge, D. González-Salgado, J. Peleteiro, E. Carballo, and G. Domarco

*Departamento de Fisica Aplicada, Universidad de Vigo, As Lagoas s/n, 32004, Ourense, Spain*

**ABSTRACT**

In this work, we studied the flux creep phenomenon over a long period [viz. about 1.5 10E7 s (i.e. for 6 months)] at liquid nitrogen temperature. For this purpose, four high-Tc ring-shaped samples were field-cooled (one sample) or ferromagnetic field-cooled (three samples) in order to induce a persistent current. The resulting current decay was measured using a Hall probe system and the results obtained revealed low relaxation rates in the ferromagnetic field-cooled samples. Also, the experimental data were found to depart from the prediction of the classical models after a long enough time. The slope of the logarithmic current decay plot exhibited an oscillatory phenomenon at ca. 2 10E5 s (about 55 hours). Oscillations vanished at ca. 4-5 10E6 s (46-55 days), after which the induced current remained stable throughout the remainder of the experiencing period.

# INTRODUCTION

One of the most comprehensively studied properties of the vortex state in high-$T_C$ superconductors are the relaxation processes of the magnetic moment $M$ and the induced current $I$ in bulk or ring shaped samples, respectively.[1,2] These processes are originated from the thermally activated hopping of fluxons between neighboring pinning positions, which is known as "flux creep". This leads to a direction drift in the magnetic flux under the action of Lorentz force that is accompanied by dissipation and gives rise to a decay in the superconducting current at $I < I_c$ (where $I_c$ is the critical current).

The earliest model used to account for this behavior was the Anderson–Kim theory of flux creep,[3–6] which assumes a logarithmic decay of the current with time that conforms to the following equation: $I(t) \sim I_c \{1 - (kT/U_0)\ln[1 + (t/t_0)]\}$, where k, $T$, $U_0$, $t$ and $t_0$ denote the Boltzmann constant, working temperature, pinning potential barrier, time of the flux creep experience and microscopic attempt time, respectively. In this context, the activation energy $U$ for flux motion is given by the expression $U \sim U_0[1 - (I/I_c)]$, where $U ? 0$ at $I \sim I_c$ and $U ? U_0$ at $I \ll I_c$. Although this theory is quite consistent with experimental facts for superconductors with strong pinning sites, it is inaccurate with collective pinning from many weak pinning sites[7–10] —which is typical of high-$T_c$ superconductors. At $I \ll I_c$, collective pinning leads to the power law dependence $U \sim U_0(I_c/I)^\mu$, µ being a numerical factor equal to or smaller than unity. Based on the foregoing, the expression for the Anderson–Kim model can be rewritten as follows: $I(t) \sim I_c\{(kT/U_0)\ln[(1 + (t/t_0)]\}^{(-1/\mu)}$. While this relation holds at $I \ll I_c$, it does not at $I \sim I_c$, where the barrier becomes zero as shown by the Anderson–Kim model. An interpolation formula accounting for the situation in between the previous two extremes could therefore be $I(t) \sim I_c\{1 - (\mu kT/U_0)\ln[1 + (t/t_0)]\}^{(-1/\mu)}$. This expression converges on that of the Anderson–Kim theory at $I \sim I_c$ and on the power law dependence at $I \ll I_c$.

Flux creep models have been developed from comprehensive studies of the temperature, initial current and pinning dependence.[11–13] Such studies have spanned periods usually not longer than $3 \times 10^4$ s and never exceeding *ca.* $3.5 \times 10^5$ s (*i.e.* about 4 days).[14,15] After this time, relaxation was assumed to continue indefinitely (Fig. 2, upper plot) in accordance with the above-described relations. In this work, we examined the flux creep phenomenon over a much longer period. For this purpose, persistent currents were induced with field cooling procedures in four ring-shaped samples and their decay was measured with Hall probes.

**EXPERIMENTAL DETAILS**

Measurements were made on two Bi-2223 (B1 and B2) and two YBCO rings (Y1 and Y2). One of the YBCO rings (Y1) was a typical top seeding melt grown YBCO with Sm seed.[16] Y2 was manufactured identically but doped with Zn to facilitate the formation of pinning sites. These bulk samples were drilled to obtain the rings. Bi-2223 polycrystalline samples were purchased from Can Superconductors, and Y1 and Y2 were obtained from the *Institut für MaterialPhysik of the University of Göttingen* and from the *Laboratorio de Baixas Temperaturas e Superconductividade of the University of Santiago de Compostela*, respectively. Table I shows the size, critical temperature and critical current of the samples. The critical temperature was taken to be the nominal value and the critical current was measured with an inductive device based on the transformed method.[17,18]

Persistent current was induced by using a classical field cooling procedure for B1 and an alternative (ferromagnetic field cooling) procedure recently developed in our laboratory for B2, Y1, and Y2.[19,20] The latter involves placing a small coil around a ferromagnetic core at room temperature with the core crossing the hole of the superconducting ring. Then, a standard DC power supply is used to generate a current that flows by the coil and magnetizes the core, thereby inducing a current in the superconducting ring that is immediately dissipated by the resistance of the sample. The whole device is cooled to a temperature below $T_C$. Finally, the DC power supply is switched off and both the inducting coil and the ferromagnetic core are removed. Therefore, a combination of the superconductor non-resistance and Faraday's law, $\oint \vec{E} \cdot \vec{dl} = -\frac{d\vec{F}}{dt}$ (where $\vec{E}$ is the electric field, $\vec{l}$ the circumference of the ring and F the magnetic flux through the hole), induces a persistent current in the superconducting ring that maintains the flux through the hole constant. The main improvement of this experimental technique upon the traditional field cooling procedure is its simplicity. A high persistent current can be induced using small coils and low probe currents.

Flux creep was studied by measuring the magnetic field decay, using a Hall probe system. This experimental technique has proved more accurate than magnetization measurements at long times; in fact, the latter are restricted to an experimental time window from 1–10 s whereas the latter afford experience times as long as $10^5$–$10^6$ s.[21] The magnitude of the current and its relaxation were calculated from the magnetic field in the sample center, using the Biot-Savart law. Experimental data for cylinder B1 were acquired using a non-cryogenic Hall probe with a time interval between measurements of 12 h the first week and 24 h afterwards. Because of the non-cryogenic nature of the Hall probe used, this was calibrated before each measurement in order to ensure adequate quality in the data. The other three samples were measured with cryogenic Hall probes, using a time interval between measurements of 60 s. Experiences were conducted over periods of approximately 4 months (sample B1) and 6 months (B2, Y1, and Y2). All experiences were performed at a working temperature of 77 K. The cryogenic probes were subjected to a stability test for about 1 month under the same experimental conditions (*i.e.* working temperature and working position) after the experiments. These additional measurements were made at zero magnetic field, using the same time interval between measurements (60 s). The purpose of this test was to ensure accuracy in the probe, with a threshold of ± 3 µT.

**RESULTS AND DISCUSSION**

*A.    Simple current decay measurements*

Figures 1–4 show the temporal variation (in logarithmic form) of the persistent current $I$ induced in the four superconducting samples. As an example, current decay for B2 (Fig. 2) was expressed in to different time scales (1 day and 6 months) to establish a comparison. These logarithmic plots are more appropriate than normal plots for the intended purpose, particularly at long times, where the asymptotic variation observed can mask some physical features. The initial current $I_0$ induced in the samples was taken to be the critical one (*i.e.* $I_0 = I_C$). Also, experimental data were normalized to the initial value, $I_0$. As can be seen, Fig. 1 exhibits poorer resolution than the other three by effect of the non-cryogenic Hall probe used and the longer time interval between measurements. The experimental data included a few non-logarithmic points at the beginning that converged on a logarithmic decay after a given interval.[8,9] Because $I_0 = I_C$, the Anderson–Kim model proved an appropriate choice for examining these curves.[3–6] In the light of this theory, the relaxation rate can be defined as $S = (1/I_C)(dI/d\ln t)$, where $S \sim kT/U_0$. $S$ and the pinning potential barrier $U_0$ can be obtained from the experimental data (see Table II). The values obtained for samples B2, Y1 and Y2 were all very similar and among the lowest reported so far, which have typically ranged from 0.01 to 0.1.[22] Note the increased $S$ value for B1 and its difference from that for B2, which was identical with it in size and chemical composition. The origin of the difference must be in the inductive technique used with the rings B2, Y1 and Y2. With the classical field cooling procedure (that used with B1), the superconducting ring is cooled with the magnetic field affecting the whole sample (ring hole and wall), so a high flux density is trapped into the wall. With our field cooling procedure, the inducting magnetic field flows mainly through the ring hole by effect of the ferromagnetic core used (Fig. 5). This results in a low flux density into the ring wall. The flux density is also related to the vortex lattice parameter of flux trapped into the ring wall, which is estimated to be $a_0 \sim 1.075(\Phi_0/B)^{1/2}$, being $B$ the applied field and $\Phi_0$ the flux quantum;[23] raising the applied field increases the flux density and decreases the vortex lattice parameter. Definitely, the relation between the density of flux trapped and the relaxation parameter $S$ comes from that for the case of a high flux density; the amount of flux required to cause dissipation is greater and their jumps result in more marked current decay (*i.e.* in an increased $S$ value).[15]

*B. Long period measurements.*

Although previous results revealed an usual behavior as regards current decay in ring-shaped superconductors, excepting the new features of the ferromagnetic field cooling procedure, measurements made at long times provide additional useful information. As can be seen from Figs 1–4 and 6, the four samples studied exhibited and oscillatory regime and a logarithmic behavior in their properties after about $2 \times 10^5$ s (approximately 55 h). Even sample B1, which was measured with a non-cryogenic Hall probe, exhibited some evidence of oscillations. Also, the oscillations exhibited a decreasing trend. This is quite logical as the current in the rings is self-maintained and the sole possible process is partial dissipation through flux creep. Therefore, an oscillation with increasing values of $I/I_0$ would be misleading. It should be noted that the oscillation does not come from an experimental threshold of measurement. As noted in the Experimental section, the cryogenic Hall probes were checked over a period of about 1 month at zero magnetic field after measurements. The experimental data thus obtained exhibited a threshold of ± 3 µT (fig. 6 (c)). Table III shows the differences (?$B_1$, ?$B_2$...) between the experimental values at the oscillation peak (denoted by (1), (2)... in the figures) and the values predicted by the Anderson–Kim model at the same time for samples B2, Y1 and Y2. These results show that, even with the weaker oscillations, its value is one order of magnitude higher than the Hall probe threshold, so the quality of the results is guaranteed.

There is one other salient feature in flux creep over long periods. Thus, as can be seen from Figs 1–4, after a time $t_e \sim 8 \times 10^6$ s (about 92 days) for sample B1 and $t_e \sim 4\text{-}5 \times 10^6$ s (about 46-55 days) for samples B2, Y1 and Y2 (Table II), the experimental data leave the previous logarithmic, oscillatory behavior and the current achieve a stabilization. Experiences were conducted over a period of approximately 4 months for sample B1, and 6 months for B2, Y1, and Y2, during which no evidence of relaxation was found. $t_e$ was obtained from the intersect of the fitted Anderson–Kim curve and the baseline of the stabilized current. Only a logarithmic plot allow us to differentiate clearly between the classical lineal decay of $I/I_0$ vs. ln $t$ and the regimen found in this work, since using a non-logarithmic plot, the stabilization regime could be masked by the asymptotic behavior of the current decay. Samples B2, Y1 and Y2 exhibited similar $t_e$ values that were all lower than that for B1. Also, the lowest value for the stabilized current, $I_t$, was that for sample B1, which exhibited the highest relaxation rate. Because $t_e$ was similar for the samples where the current was induced using the same procedure and clearly different from the value for the other, the differential density of the trapped flux again appears to be an important factor. Thus, the high flux density results in a high $S$ values and appears to confirm the occurrence of relaxation after a long time, with a low value for $I_t$. The value of the stabilized current, $I_t$, for the different samples is also shown in Table II.

Although the primary focus of this work was the experimental behavior of flux creep over long periods, we also suggest a physical explanation for the oscillation and stabilization phenomena observed. In a flux experience, thermally activated trapped flux is led by the Lorentz force to the outer or inner side of the superconducting ring wall (*i.e.* in the radial direction). As the experience progresses, a large amount of flux accumulates on one of the wall sides at the expense of the other. Therefore, the amount of flux moving from one side to the other also decreases, and a logarithmic current decay (the Anderson model) is observed as a result. Under these conditions, the oscillations observed after about 55 h can be due to the collection of flux at intermediate strong pinning sites (*e.g.* YBCO-211 inclusions). In this situation, the slope of the current decay decreases while the summation of the Lorentz force and thermal fluctuations does not exceed the pinning force. As the experience develops, more flux reaches the pinning sites and as soon as the pinning force is surpassed, the magnetic flux jumps. The dissipation regime is thus reactivated and the slope of the decay increases again. Clearly, the slope decreases but is never zero under this regime, as a certain amount of flux never interacts with these pinning sites and can flow freely.

The explanation for the end of the relaxation phenomenon must lie in the end of flux jumping. As noted earlier, as experiences progress, the magnetic flux is displaced to the pinning positions on one side of the superconducting wall. The boundary of the superconducting wall with its neighboring positions probably becomes a very strong potential barrier that pins the magnetic flux, thus enabling its motion and bringing the vortex lattice into balance. As a result, the current stabilizes in a natural manner.

Although our results illustrate the experimental behavior of flux creep after a long time, much research remains to be done in this field—specific work in this direction is currently under way in our laboratory. Thus, it would be interesting to examine the theoretical aspects of the oscillation and stabilization phenomena, which appear to raise no experimental doubts. It would also be interesting to check the relationship between the stabilization time and the oscillation at different types of pinning and with materials of different size but the same composition.

**CONCLUSION**

We made current decay measurements of ring-shaped high-$T_C$ samples at long times using a Hall probe system at 77 K. The lowest relaxation rates were exhibited by the samples where the current was induced using the ferromagnetic field cooling procedure; the results were consistent with the lowest values reported to date. Also, measurements at long times revealed the presence of oscillations in the current decay after *ca.* $2 \times 10^5$ s; this behavior was maintained up to about $4\text{-}5 \times 10^6$ s, where relaxation seemingly ends and the current stabilizes throughout the remainder of the measurement time.

**ACKNOWLEDGEMENTS**

The authors wish to thank M.T. González and the research group headed by Professor Félix Vidal for their helpful collaboration.

**TableI:** Technical characteristics of the superconducting rings used in this work. $T_C$, $I_C$ are the critical temperature and current, respectively.

| Sample | $T_C$ (K) | $I_C$ (A) | Inner diameter (mm) | Height (mm) | Width (mm) |
|--------|-----------|-----------|---------------------|-------------|------------|
| B1     | 108       | 62        | 10                  | 10          | 1          |
| B2     | 108       | 54        | 10                  | 10          | 1          |
| Y1     | 93        | 290       | 10                  | 14          | 4          |
| Y2     | 93        | 117       | 10                  | 15          | 10         |

**Table II:** Initial current induced in the samples $I_0$, relaxation rate $S$, pinning potential barrier $U_0$, stabilization time $t_e$, and the stabilized current at this time $I_t$.

| Sample | $I_0$ (A) | $S$   | $U_0$ (eV) | $\sim t_e$ (s) $\times 10^6$ | $I_t$ (A)  |
|--------|-----------|-------|------------|------------------------------|------------|
| B1     | 62        | 0.059 | 0.1        | 8                            | 0.68 $I_0$ |
| B2     | 54        | 0.013 | 0.5        | 4                            | 0.87 $I_0$ |
| Y1     | 290       | 0.010 | 0.6        | 4- 5                         | 0.93 $I_0$ |
| Y2     | 117       | 0.016 | 0.4        | 4 - 5                        | 0.82 $I_0$ |

**Table III:** Differences between the Anderson model and the experimental values of the oscillation peak at the same time for the samples B2, Y1, and Y2. $\Delta B_1$, $\Delta B_2$, ... denote the mentioned difference for the peaks numbered as (1), (2), ..., respectively.

| B2 | | Y1 | | | Y2 | |
|---|---|---|---|---|---|---|
| $\Delta B_1$ (mT) | $\Delta B_2$ (mT) | $\Delta B_1$ (mT) | $\Delta B_2$ (mT) | $\Delta B_3$ (mT) | $\Delta B_1$ (mT) | $\Delta B_2$ (mT) |
| 68 | 34 | 182 | 553 | 327 | 185 | 238 |

**Figure Captions.**

Figure 1. Logarithmic plot of normalized current vs. time for the B1 ring. (◊) Experimental data, (−) extrapolated Anderson fitting, and (---) stabilized base line.

Figure 2. Logarithmic plot of normalized current vs. time for the B2 ring. (◊) Experimental data, (−) extrapolated Anderson fitting, and (---) stabilized base line. Two oscillation regions are denoted by (1) and (2). The upper plot of the figure depicts current decay during a day.

Figure 3. Logarithmic plot of normalized current vs. time for the Y1 ring. (◊) Experimental data, (−) extrapolated Anderson fitting, and (---) stabilized base line. Three oscillation regions are denoted by (1), (2), and (3).

Figure 4. Logarithmic plot of normalized current vs. time for the Y2 ring. (◊) Experimental data, (−) extrapolated Anderson fitting, and (---) stabilized base line. Two oscillation regions are denoted by (1) and (2).

Figure 5. Scheme about the differences between the ferromagnetic (A1 - A2) and the traditional (B1 – B2) field cooling. In the ferromagnetic field cooling procedure, the ferromagnetic core through the sample's hole guides the inducting field $\vec{B}_1$ (A1). After the induction, the trapped flux $f$ into the sample's wall (A2) is a result of the small leaks of the ferromagnetic core. The bases of field cooling procedure is the cooling with a magnetic field $\vec{B}_1$ acting to the ring hole and wall (B1). The induction produces a higher density of trapped flux $f$ (B2).

Figure 6. (a) Detail of the oscillations for the ring Y1. (◊) Experimental data and (−) extrapolated Anderson fitting. $\Delta B_1$, $\Delta B_2$, and $\Delta B_3$ denote the difference between the Anderson value and the experimental value for the peaks 1, 2, and 3, respectively. (b) Plot of the Hall probe test at zero field at the same scale of (a). (c) Noise of one cryogenic Hall probes.

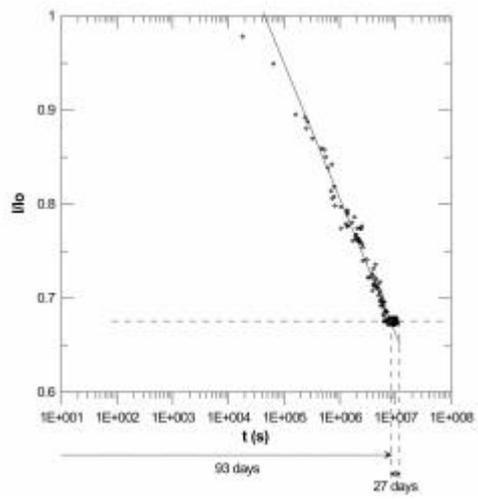

**Figure 1**

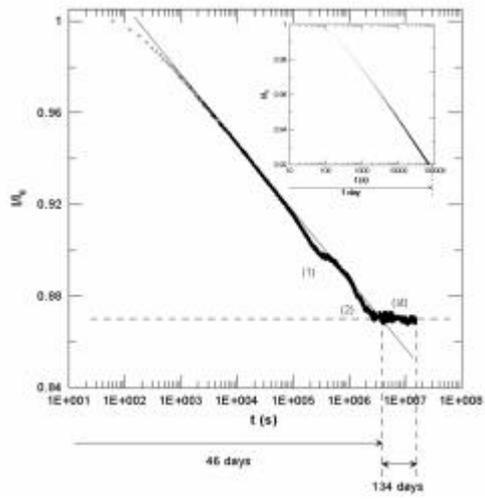

**Figure 2**

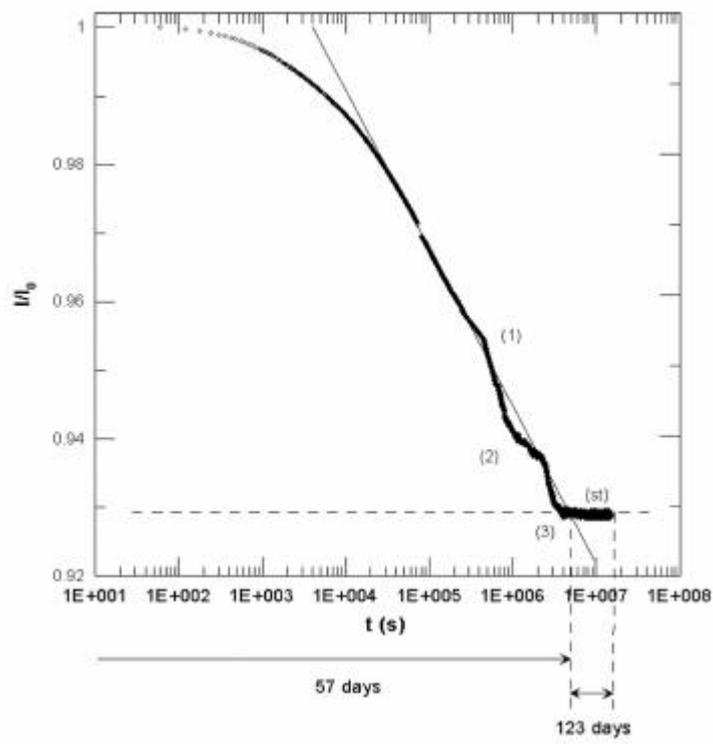

**Figure 3**

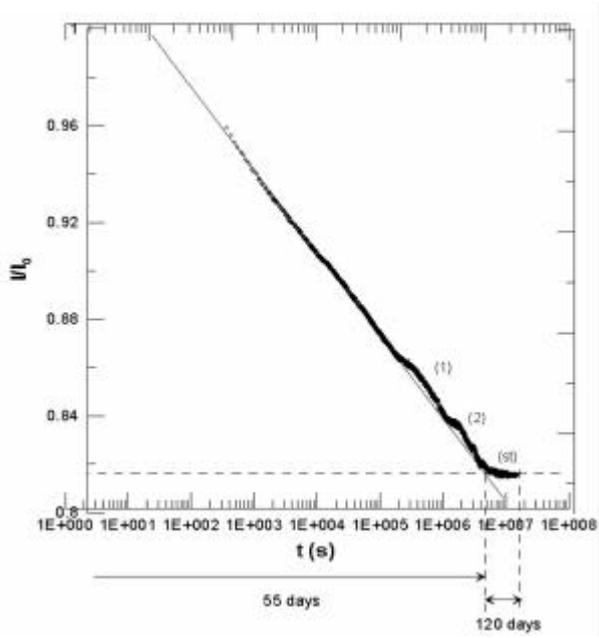

**Figure 4**

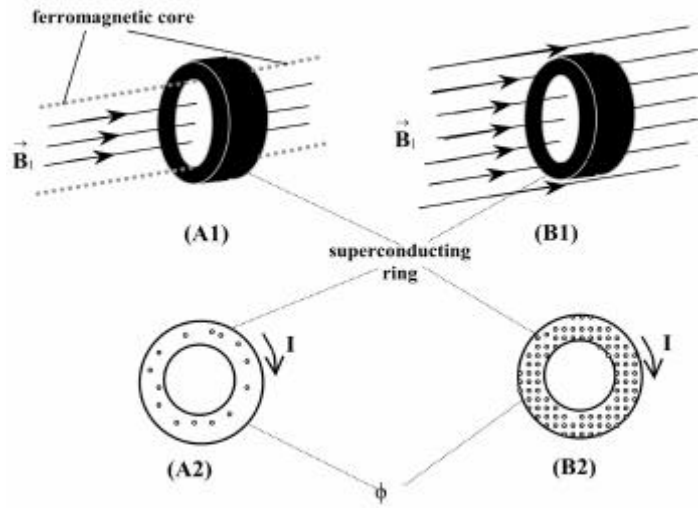

**Figure 5**

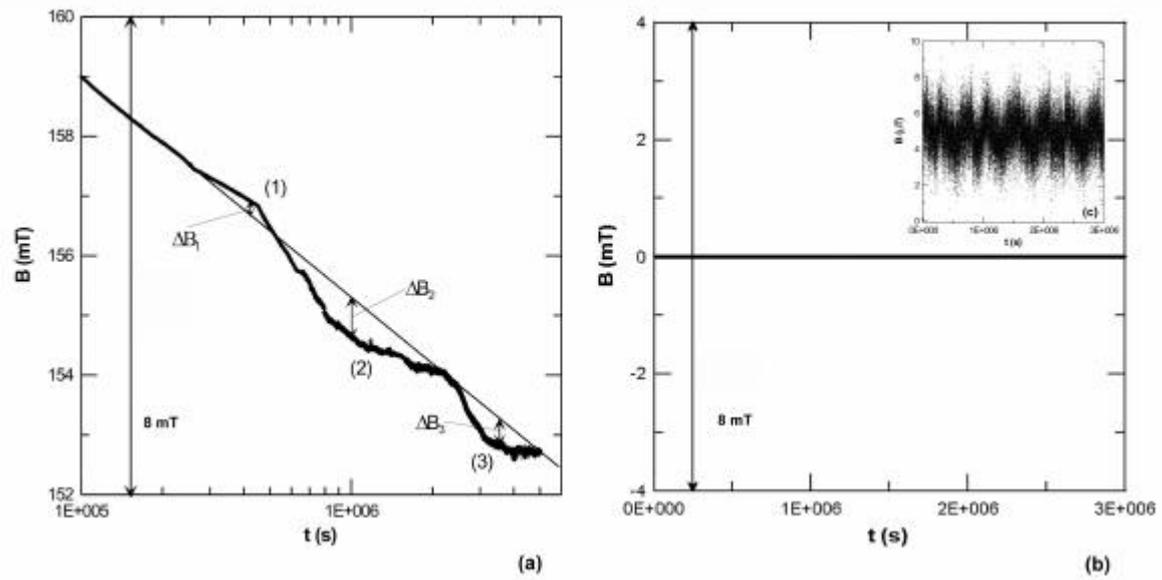

**Figura 6**